\RequirePackage{fix-cm}
\documentclass[smallextended]{svjour3}       
\smartqed  
\usepackage{graphicx}
\begin{document}

\title{Re-Examination of Simple Kaluza-Klein Cosmologies}

\author{Darrell Jackson}


\institute{D. Jackson \at
              Applied Physics Laboratory \\
              University of Washington \\
              Seattle, WA USA \\
              Tel.: 206-909-9678\\
              Fax: 206-543-6785\\
              \email{drj12@uw.edu}           
}

\date{Received: date / Accepted: date}

\maketitle

\begin{abstract}
Simple cosmological models based upon five-dimensional Kaluza-Klein relativity
are re-examined and interesting properties are indicated. These models are special
cases of those obtained by Davidson et al. and Mann and Vincent, specifically, 
those with vanishing cosmological constant. 
No five-dimensional sources are present: the solutions
are Ricci-flat in five dimensions. Electromagnetic degrees of freedom are assumed
not to be excited, consequently the four-dimensional stress-energy tensor induced
by dimensional reduction is entirely due to the
scalar field, obeying the radiation equation of state.
For the three choices of curvature index, the dependence of
the scale factor on cosmic time corresponds to, for
$k = -1$ either a bounce or big bang, for $k = 0$ a big bang,
and for $k = 1$ a big bang followed by collapse. The Kretschmann scalar is 
proportional to the square of acceleration and approaches zero for $k = -1$
and $k = 0$, while revealing true singularities in some cases.
Only one of the models exhibits the compactification behavior hoped for in a realistic K-K model:
shrinking of the circumference of fifth dimension in the earliest times followed by an extended 
period of stability. 

\end{abstract}

\section{Introduction}
\label{Intro}
Five-dimensional Kaluza-Klein (K-K) General Relativity has been used in the development of numerous
cosmological models. Of recent work, some has adhered to the spirit of the orginal K-K
paradigm, with compact extra dimension \cite{Mukhopadhyay_et_al:16} and some has treated the extra dimension
as non-compact, a parameter defining a 4D universe embedded in a flat 5D space \cite{Wesson:06}, \cite {Wesson:15}.
The variety in K-K cosmological models stems not only from the
choice between compactness or its lack, but also in
presence or absence of sources in 5D, presence or absence of a cosmological constant, and
inclusion of more or fewer of the degrees of freedom allowed by the 5D metric.
If one seeks the ``purest" form of K-K cosmology, one guided by Einstein's hope for complete
geometrization of physics, models without 5D sources and cosmological constants would be preferred.
This article considers models of this sort, further simplified by the assumption of
the absence of electromagnetic fields. 
Although compactness of the fifth dimension is not demanded by the analysis, the motivation for the present
work hinges upon Klein's assumption of closure in this dimension.

The restrictions noted above lead to models that are special cases of those introduced  in \cite{Davidson:85}
and \cite{Mann_Vincent:85}: those with vanishing cosmological constant. These solutions will be examined 
more closely than in prior work in order to display the time evolution of key cosmological parameters
and to identify properties that must be altered in developing more realistic models.

\section{Field Equations}
\label{field_eqs}
Following the dictum that the Einstein equations in 5D should have no source terms on the right-hand side,
the cosmological models to be discussed obey
\begin{equation}
\label{Eq:field_eq_5D}
^{(5)}R_{AB} = 0~.
\end{equation}
The superscript (5) is used here to distinguish the 5D Ricci tensor from its 4D counterpart,
and no superscript is used to denote 4D tensors.
The indices $A$ and $B$ range over all five coordinates.
Dimensional reduction gives, in place of (\ref{Eq:field_eq_5D}), Einstein's 4D equation
(in units where $G = 1$ and $c = 1$)
\begin{equation}
\label{Eq:field_eq_Einstein}
G_{\mu \nu} = 8 \pi T_{\mu \nu}~,
\end{equation}
with
\begin{equation}
\label{Eq:stress_energy}
8 \pi T_{\mu \nu} = \psi_{;\mu \nu} + \psi_{,\mu} \psi_{,\nu}~,
\end{equation}
where the scalar field $\psi$ is related to the fifth metric component as follows:
\begin{equation}
\label{Eq:psi}
g_{LL}(t) = e^{2 \psi}~. 
\end{equation}
Greek indices range over the usual four coordinates, and the fifth coordinate is labeled $L$.
The electromagnetic vector potential, $g_{\mu L}$, is assumed to vanish, in which case the 
equation of motion for the scalar field is
\begin{equation}
\label{Eq:psi_wave_eq}
\psi_{;\mu}^{~~\mu} + \psi_{,\mu} \psi_{;}^{~\mu} = 0~.
\end{equation}
This equation guarantees that the stress-energy tensor is conserved, that is, has vanishing divergence.
It also leads to the conclusion that the stress-energy tensor is traceless, which, in turn, implies that
the 4D curvature scalar $R = R_{\mu}^{\mu}$ vanishes, reducing (\ref{Eq:field_eq_Einstein}) to
\begin{equation}
\label{Eq:field_eq_Einstein2}
R_{\mu \nu} = 8 \pi T_{\mu \nu}~.
\end{equation}

To address the cosmological problem, the metric will be taken as
the Friedmann-Robertson-Walker (FRW) metric extended in the simplest way to five dimensions,
with line element
\begin{equation}
\label{Eq:line_element_5D}
ds^2 = -dt^2 + a^2(t)[\frac{dr^2}{1-kr^2} + r^2 d\Omega^2] + g_{LL}(t) dL^2~. 
\end{equation}
In accordance with the cosmological principle, the fifth metric component
is assumed to depend only upon cosmic time. The curvature index $k$ may take on values -1, 0 , and 1.

With this metric, (\ref{Eq:field_eq_Einstein}) gives the usual equations of motion for the scale factor
\begin{equation}
\label{Eq:a_dot}
{\dot a}^2 =  8 \pi \rho a^2 /3 - k~, 
\end{equation}
\begin{equation}
\label{Eq:a_ddot}
{\ddot a} = - 8 \pi (\rho + 3 p) a/6~, 
\end{equation}
where the dot indicates a derivative with respect to $t$. The density is
\begin{equation}
\label{Eq:density}
8 \pi \rho = {\ddot \psi} + {\dot \psi}^2~, 
\end{equation}
and the pressure is
\begin{equation}
\label{Eq:pressure}
8 \pi p = - {\dot a} {\dot \psi}/a~. 
\end{equation}
Equation (\ref{Eq:psi_wave_eq}) becomes
\begin{equation}
\label{Eq:psi_wave_eq2}
{\ddot \psi}+ 3 \frac{\dot a} a {\dot \psi} + {\dot \psi}^2 = 0~.
\end{equation}
Using this, the density (\ref{Eq:density}) is seen to satisfy the equation of state for radiation,
$p = \rho / 3$.

As a first step toward finding solutions, integrate (\ref{Eq:psi_wave_eq2}) to obtain
\begin{equation}
\label{Eq:psi_wave_eq_int2}
{\dot \psi}a^3 e^{\psi} = \alpha~,
\end{equation}
where $\alpha$ is an integration constant to be determined.
Next, combining (\ref{Eq:a_ddot}) - (\ref{Eq:psi_wave_eq2}) yields
\begin{equation}
\label{Eq:a_ddot2}
{\ddot a} = {\dot a} {\dot \psi}~.
\end{equation}
This can be integrated to obtain
\begin{equation}
\label{Eq:exp_psi}
e^{\psi} = \beta {\dot a}~,
\end{equation}
where $\beta$ is another integration constant to be determined.
This provides a simple connection between the fifth metric component, $\exp(2\psi)$, and
the rate of increase of the scale factor. 
If (\ref{Eq:a_dot}), (\ref{Eq:density}), (\ref{Eq:psi_wave_eq2}) and (\ref{Eq:a_ddot2}) 
are combined, a second-order equation for the scale factor can be found.
\begin{equation}
\label{Eq:a_ddot3}
{\dot a}^2 + a {\ddot a} + k = 0~.
\end{equation}
The second derivative can be eliminated by differentiating (\ref{Eq:exp_psi}) and substituting
in (\ref{Eq:psi_wave_eq_int2}). The result is the following first-order equation for the scale factor,
\begin{equation}
\label{Eq:a_dot2}
{\dot a}^2 = - \frac{\alpha}{\beta a^2} -k~.
\end{equation}
Solutions of (\ref{Eq:a_dot2}) for specific choices of $k$ will be given in Sect. \ref{sec:solutions}.

\section{Curvature}
\label{sec:curvature}
As has been shown above, the 4D curvature scalar vanishes. As the 5D Ricci tensor vanishes, 
so must the 5D curvature scalar vanish. To show that the 5D space is not flat, as it is in
some K-K cosmologies, and to reveal singularities, the 5D Kretschmann scalar will be evaluated. It is defined as
\begin{equation}
\label{Eq:kretschmann}
K = R^{abcd}R_{abcd}~,
\end{equation}
and is found to be
\begin{equation}
\label{Eq:kretschmann2}
K = 12 {\big [}\frac {{\ddot a}^2}{a^2} + \frac{(k+{\dot a}^2)^2}{a^4} + \frac {{\dot a}^2 {\dot \psi}^2}{a^2} + \frac 1 3 ({\ddot \psi} + {\dot \psi}^2)^2 {\big ]}~.
\end{equation}
The Kretschmann scalar is uniform over all five spatial dimensions and dependent only on time. The first two terms in 
(\ref{Eq:kretschmann2})
give the Kretschmann scalar for the 4D FRW metric, and the remaining terms result from adding the fifth dimension. 
Using the equations of motion,
\begin{equation}
\label{Eq:kretschmann3}
K = 72 {\big (} \frac {\ddot a} a {\big )}^2~.
\end{equation}
This holds for all choices of $k$ and shows that the model universe will tend toward 5D flatness only if the magnitude
of acceleration tends toward zero as $t$ increases. The Kretschmann scalar for the 4D sub-space is of the same form as
(\ref{Eq:kretschmann3}), except the factor 72 is replaced by 24.

\section{Density Parameter}
\label{sec:Density}
It will be convenient to evaluate the density parameter for the solutions to be given later. This parameter is
\begin{equation}
\label{Eq:Omega}
\Omega = \frac {\rho}{\rho_c}~,
\end{equation}
where
\begin{equation}
\label{Eq:critical_density}
\rho_c = \frac{3 H^2}{8 \pi}
\end{equation}
is the critical density and
\begin{equation}
\label{Eq:hubble}
H = \frac {\dot a} a
\end{equation}
is the Hubble parameter. Using (\ref{Eq:density}) and (\ref{Eq:a_ddot2}),
\begin{equation}
\label{Eq:Omega+equals_q}
\Omega = q~,
\end{equation}
where 
\begin{equation}
\label{Eq:q}
q = -\frac {a {\ddot a}}{{\dot a}^2}
\end{equation}
is the deceleration parameter. The equality expressed in (\ref{Eq:Omega+equals_q}) 
is interesting, as it is independent of $k$. Even so,
$q$ and $\Omega$ depend on the choice of $k$:
\begin{equation}
\label{Eq:critical_density2}
\Omega = 1 +\frac k {{\dot a}^2}~.
\end{equation}

\section{Solutions}
\label{sec:solutions}
Solutions for the various choices of $k$ will be examined in this section. 
These solutions are equivalent to the $\Lambda = 0$ solutions
of \cite{Davidson:85} and \cite{Mann_Vincent:85}. The Kretschmann scalar will reveal true singularities, 
the density parameter will provide a measure of departure from realistic cosmologies as well
as quantifying acceleration, and $g_{LL}$ will indicate tendency toward or away from compactification 
as time increases.

\subsection{Negative $k$}
\label{sec:k_minus_1}
For $k = -1$, (\ref{Eq:a_dot2}) has solution
\begin{equation}
\label{Eq:t_a_k_minus_1}
t = \sqrt{a^2-\alpha / \beta} + t_0~,
\end{equation}
where $t_0$ is an integration constant. There are two cases to consider: Case 1 where $\alpha / \beta < 0$
and Case 2 where $\alpha / \beta > 0$. 

\begin{figure}[h] 
  \centering
\includegraphics[width=5.0in,keepaspectratio] {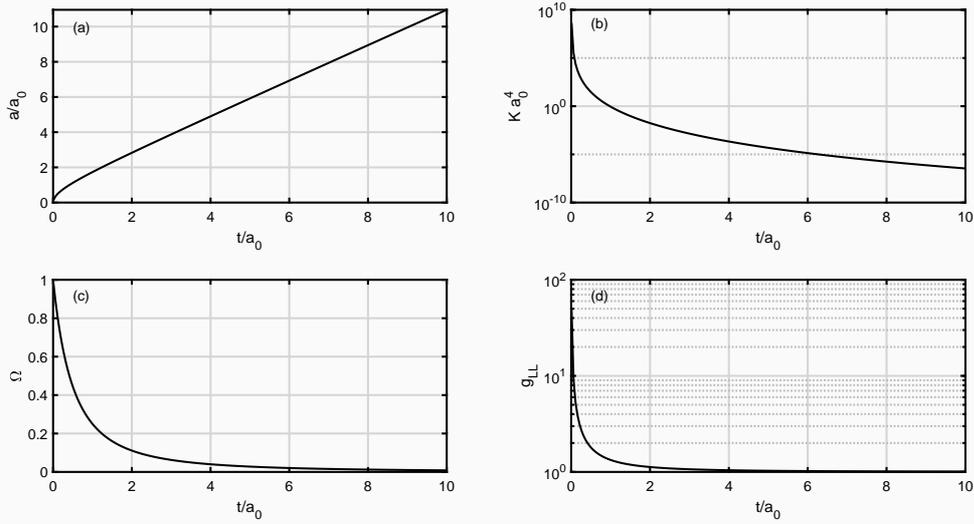}
\caption{Time dependence of cosmological parameters for $k = -1$, Case 1. (a) scale factor,
(b) Kretschmann scalar, (c) Density parameter, equal to deceleration parameter, 
(d) metric component for fifth coordinate.}
\label{fig:1}       
\end{figure}

For Case 1,
\begin{equation}
\label{Eq:a_k_minus_1_case_1}
a = \sqrt{t^2 + 2 a_0 t}~.
\end{equation}
Here, integration constants have been chosen so that $a = 0$ at $t = 0$, with $a_0 = \sqrt{- \alpha / \beta}$. 
As seen in Fig. \ref{fig:1},
the scale factor is an hyperbola in $t$, convex upward,  with a big bang at $t = 0$ and with asymptote $a \rightarrow t$. Continuing with Case 1,
\begin{equation}
\label{Eq:K_k_minus_1_case_1}
K = 72 \frac {a_0^4}{(t^2 + 2 a_0 t)^4}~,
\end{equation}
\begin{equation}
\label{Eq:Omega_k_minus_1_case_1}
\Omega = \frac {a_0^2}{(t + a_0)^2}~,
\end{equation}
\begin{equation}
\label{Eq:K_k_minus_1_case_1}
g_{LL} = \frac {(t+a_0)^2}{t^2 + 2 a_0 t}~.
\end{equation}
The integration constant $\beta$ has been chosen so that $g_{LL}$ asymptotes to unity.
The Kretschmann scalar diverges as $t$ approaches zero, confirming the presence of a true
singularity at $t = 0$. While the density parameter is perfectly tuned to unity at this instant,
it approaches zero as time progresses. This shows that deceleration also approaches zero in this limit.
The fifth metric component is infinite at $t = 0$ and then compactifies, approaching
unity as $t$ increases.

\begin{figure}[h] 
  \centering
\includegraphics[width=5.0in,keepaspectratio] {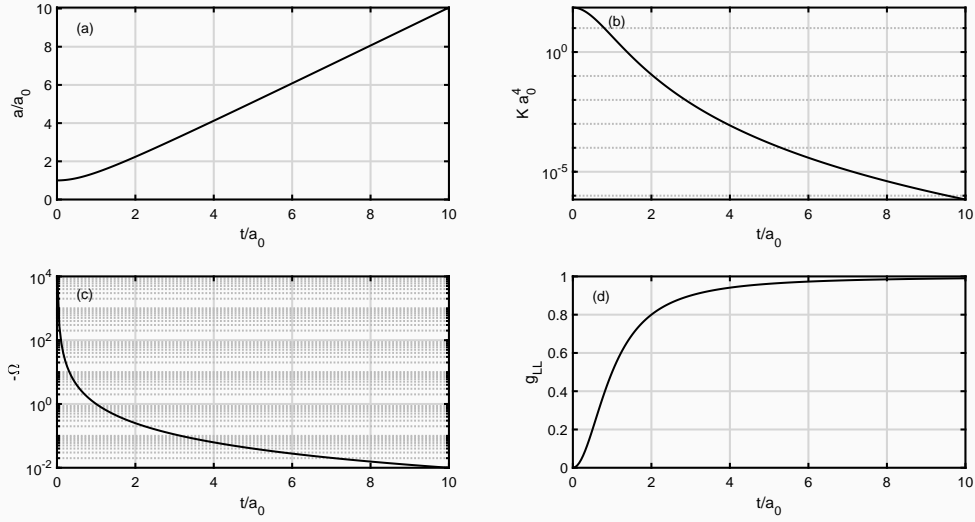}
\caption{Time dependence of cosmological parameters for $k = -1$, Case 2. (a) scale factor,
(b) Kretschmann scalar, (c) Negative of density parameter, equal to negative of deceleration parameter, 
showing acceleration due to bounce, (d) metric component for fifth coordinate.}
\label{fig:2}       
\end{figure}

For Case 2 when $k = -1$,
\begin{equation}
\label{Eq:a_k_minus_1_case_2}
a = \sqrt{t^2 + a_0^2}~.
\end{equation}
\begin{equation}
\label{Eq:K_k_minus_1_case_2}
K = 72 \frac {a_0^4}{(t^2 + a_0^2)^4}~,
\end{equation}
\begin{equation}
\label{Eq:Omega_k_minus_1_case_2}
\Omega = -\frac {a_0^2}{t^2}~,
\end{equation}
\begin{equation}
\label{Eq:K_k_minus_1_case_2}
g_{LL} = \frac {t^2}{t^2 + a_0^2}~.
\end{equation}
Referring to Fig. \ref{fig:2},
the scale factor is an hyperbola in $t$, convex downward,  with a bounce at $t = 0$,
and with asymptote $a \rightarrow t$. 
The Kretschmann scalar is finite for all times, approaching zero as time increases. 
The density parameter is negative, infinite at $t=0$, indicating infinite acceleration
at that moment.
The fifth metric component shows growth with time, starting from zero and
approaching a constant asymptote, a sort of reverse compactification.

\subsection{Vanishing $k$}
\label{sec:k_0}
In addition to being a special case in \cite{Davidson:85} and \cite{Mann_Vincent:85},
the $k = 0$ solution has been described in \cite{Wesson:92} and \cite{Wesson:99} 
and has been derived as a 5D Kasner solution in \cite{Chodos_Detweiler:87}. 
For $k = 0$, (\ref{Eq:a_dot2}) has solution
\begin{equation}
\label{Eq:a_k_0}
a = \sqrt{a_0t}~.
\end{equation}
From this,
\begin{equation}
\label{Eq:K_k_0}
K =  \frac 9{4 t^4}~,
\end{equation}
\begin{equation}
\label{Eq:Omega_k_0}
\Omega = 1~,
\end{equation}
\begin{equation}
\label{Eq:K_k_0}
g_{LL} = \frac {a_0}{4 t}~.
\end{equation}
The scale factor is a parabola as a function of time,
and the Kretschmann scalar diverges as $t$ approaches zero, confirming the presence of a true
singularity at $t = 0$. As expected for $k = 0$, the density parameter and deceleration parameter 
remain at unity for all time.
The fifth metric component is infinite at $t = 0$ and then compactifies in an extreme way, approaching
zero as $t$ increases. In a realistic model, this metric component should approach a non-zero constant.

\begin{figure}[h] 
  \centering
\includegraphics[width=5.0in,keepaspectratio] {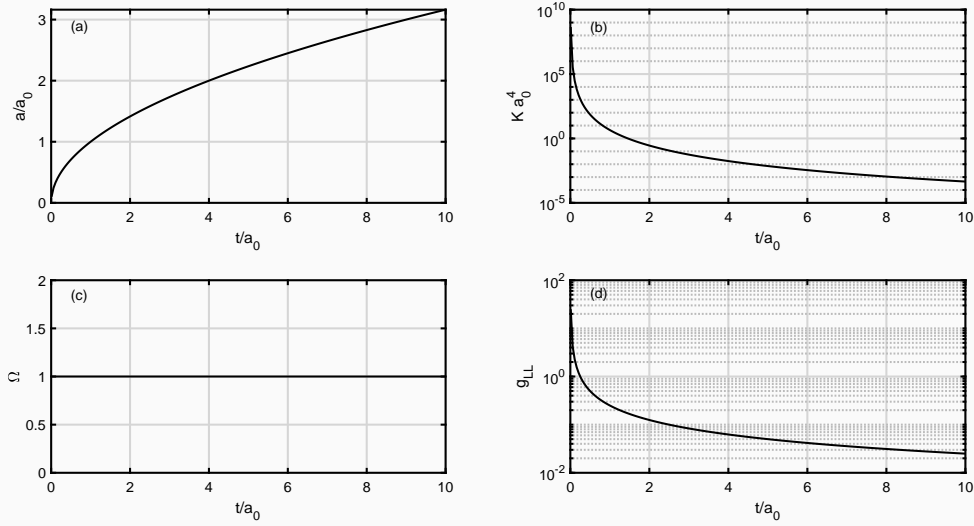}
\caption{Time dependence of cosmological parameters for $k = 0$. (a) scale factor,
(b) Kretschmann scalar, (c) Density parameter, equal to deceleration parameter, 
(d) metric component for fifth coordinate.}
\label{fig:3}       
\end{figure}

\begin{figure}[h] 
  \centering
\includegraphics[width=5.0in,keepaspectratio] {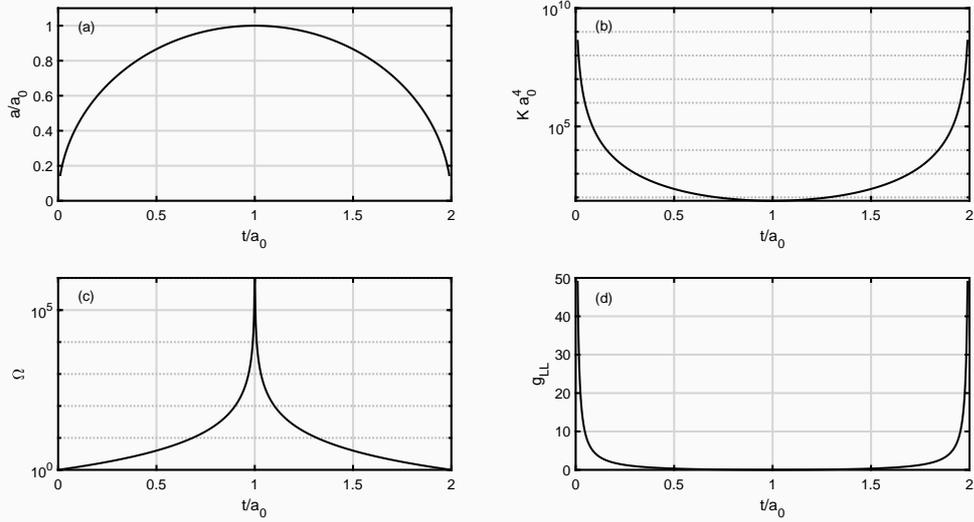}
\caption{Time dependence of cosmological parameters for $k = 1$. (a) scale factor,
(b) Kretschmann scalar, (c) Density parameter, equal to deceleration parameter, 
(d) metric component for fifth coordinate.}
\label{fig:4}       
\end{figure}

\subsection{Positive $k$}
\label{sec:k_1}
For $k = 1$, (\ref{Eq:a_dot2}) has solution
\begin{equation}
\label{Eq:a_k_1}
a = \sqrt{2 t a_0 - t^2}~,
\end{equation}
which gives
\begin{equation}
\label{Eq:K_k_1}
K =  72 \frac {a_0^4}{(2 t a_0 - t^2)^4}~,
\end{equation}
\begin{equation}
\label{Eq:Omega_k_1}
\Omega = \frac{a_0^2}{(a_0 - t)^2}~,
\end{equation}
\begin{equation}
\label{Eq:K_k_1}
g_{LL} = \frac {(a_0-t)^2}{2 t a_0 - t^2}~.
\end{equation}
The scale factor is a semicircle as a function of $t$, and $0 < t < 2 a_0$.
The Kretschmann scalar diverges both at the big bang at $t=0$ and at the completion of
collapse at $t = 2 a_0$. The density parameter and deceleration parameter are unity at
$t= 0$ and $t = 2 a_0$, and spike to infinity at $t = a_0$ when the scale factor has reached its maximum.
The fifth metric component is infinite at $t= 0$ and $t = 2 a_0$ and vanishes at $t = a_0$.
This behavior is at odds with the hope for a protracted period of stable $g_{LL}$.

\section{Discussion}
\label{sec:discussion}
The simple models discussed here lead to a simple result: in every case the
dependence of the scale factor on cosmic time is a conic section.
One cannot expect these models to be
realistic, as matter and electromagnetic radiation have been excluded.
Nevertheless, it seems useful to catalog both their good and bad 
properties.  The first of the $k = -1$ cases is attractive in exhibiting
compactification and asymptotic stability of the fifth dimension, but unattractive in the
decay of the density parameter with time. 
The second case shows acceleration, but not of the
sort associated with the cosmological constant. Also, it does not show compactification.
The $k = 0$ model is realistic in its constancy of the density parameter, but does not
show acceptable compactification. There is little to recommend the $k = 1$ model, which is
unrealistic in its density parameter and has unsuitable compactification.

While the simple models examined here show good and bad properties, it can be argued that they
are necessary components of a truly geometrical approach to Kaluza-Klein cosmology.
Addition of matter will alter the behavior of these models, perhaps for the better. 
In the Einstein-Kaluza-Klein spirit, matter would not be added phenomenologically, 
but would emerge as a solution
of the empty five-space field equations, namely, the vanishing of the 5D Ricci tensor.
Such matter would not only shape space-time, it would be altered by 5D space-time,
as particle masses would depend on the circumference of the fifth dimension. 
The presently observed constancy of masses and physical parameters places a strong 
constraint on the asymptotic behavior of the circumference in a realistic model.


%
%

\end{document}